\documentclass[conference]{IEEEtran}
\IEEEoverridecommandlockouts
\usepackage{amsmath,graphicx}
\usepackage{amsmath,amssymb,amsfonts}
\usepackage{cite, soul}
\usepackage{url}
\usepackage{bm}
\usepackage{comment}

\usepackage[usenames,dvipsnames]{color}

\title{A Hybrid Joint Source-Channel Coding Scheme for Mobile Multi-hop Networks}
\author{Chenghong Bian, Yulin Shao, Deniz G{\"u}nd{\"u}z\\
Department of Electrical and Electronic Engineering, Imperial College London.\\
\{c.bian22, y.shao, d.gunduz\}@imperial.ac.uk
}

\begin{document}

\maketitle
\begin{abstract}
We propose a novel hybrid joint source-channel coding (JSCC) scheme for robust image transmission over multi-hop networks. In the considered scenario, a mobile user wants to deliver an image to its destination over a mobile cellular network. We assume a practical setting, where the links between the nodes belonging to the mobile core network  are stable and of high quality, while the link between the mobile user and the first node (e.g., the access point) is potentially time-varying with poorer quality. In recent years, neural network based JSCC schemes (called DeepJSCC) have emerged as promising solutions to overcome the limitations of separation-based fully digital schemes. However, relying on analog transmission, DeepJSCC suffers from noise accumulation over multi-hop networks. Moreover, most of the hops within the mobile core network may be high-capacity wireless connections, calling for digital approaches. To this end, we propose a hybrid solution, where DeepJSCC is adopted for the first hop, while the received signal at the first relay is digitally compressed and forwarded through the mobile core network.  We show through numerical simulations that the proposed scheme is able to outperform both the fully analog and fully digital schemes. Thanks to DeepJSCC it can avoid the cliff effect over the first hop, while also avoiding noise forwarding over the mobile core network thank to digital transmission. We believe this work paves the way for the practical deployment of DeepJSCC solutions in 6G and future wireless networks.

\end{abstract}
%


\section{Introduction}
\label{sec:intro}

\begin{figure}
\centering
\includegraphics[width=0.9\columnwidth]{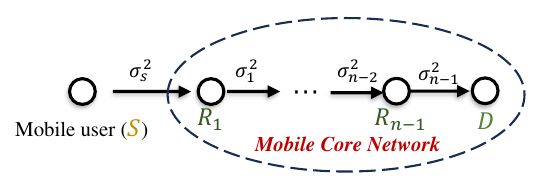}
\caption{Basic structure of the multi-hop communication network where the mobile user $\mathrm{S}$ communicates with the destination $\mathrm{D}$ with the aid of the relays $\{\mathrm{R}_1, \ldots, \mathrm{R}_{n-1}\}$ located in a {mobile core network with high-capacity wireless connections}.}
\label{fig:fig_system}
\end{figure}

Communication over multi-hop networks, where a source node delivers a message to a destination node with the aid of multiple relays has been extensively studied over the past decades. Unlike the cooperative relay channel \cite{relay_capacity, relay_capacity2}, in the multi-hop setting, each node can only communicate with its subsequent node, and there is no direct link between nodes that are two hops away \cite{Gunduz:TWC:10}. We consider a practical multi-hop scenario as shown in Fig. \ref{fig:fig_system}, where the destination and the relay nodes are located in the mobile core network\footnote{We use the term `mobile' core network to distinguish from the 5G core network which assumes wired connections.} with more stable links with higher qualities, while the first hop is a mobile wireless link with versatile channel conditions. {The considered scenario resembles the Integrated Access and Backhaul (IAB) \cite{IAB} setup in 5G, where digital coded communication scheme is employed among wireless hops in the mobile core network.} Furthermore, we are interested in the joint source channel coding (JSCC) problem, studying the end-to-end image delivery performance over this multi-hop network.

In recent years, deep neural networks (DNNs) have been successfully applied to the design of JSCC schemes over wireless channels. First proposed in \cite{deepjscc}, it has been shown that the so-called DeepJSCC approach can outperform digital alternatives employing state-of-the-art image compression techniques protected by near-optimal channel codes, and provide graceful degradation with weakening channel quality. {It is soon applied to different point-to-point (p2p) wireless channels such as multi-path fading and MIMO channels \cite{jsccofdm, ST_JSCC}.} More recently, DeepJSCC techniques have also been extended to multi-user scenarios. For instance, \cite{relay_jscc, mac_jscc, broadcast} show the superior performance of the DeepJSCC in the three node cooperative relay, the multiple access, and broadcast channels, respectively. The main advantage of DeepJSCC stems from the continuous-amplitude transmissions, whose estimation quality at the receiver depends continuously on the channel quality. This is unlike digital schemes, which are either decoded reliably, or the information is lost completely, resulting in a cliff-effect in the performance. However, this advantage of DeepJSCC becomes a curse in multi-hop networks due to noise accumulation. While digital transmission achieve the same performance independent of the number of hops, the performance of DeepJSCC degrades gradually with the number of hops. 
Recently, authors in \cite{semantic_multi-hop1, semantic_multi-hop2} provide initial investigation for JSCC over multi-hop networks. In particular, \cite{semantic_multi-hop1} introduces a term named semantic similarity and optimizes the neural networks to preserve the similarity over the hops, whereas \cite{semantic_multi-hop2} proposes a recursive training methodology to mitigate the noise accumulation problem. Though some performance improvements are attained compared with the naively trained scheme, \cite{semantic_multi-hop2} still degrades when the number of hops becomes larger. 

In this paper, to overcome the shortcomings of the aforementioned fully analog and fully digital schemes, we propose a hybrid solution, named JSC, where DeepJSCC is adopted for the first hop with versatile quality to avoid the cliff effect, while the information is forwarded digitally over the remaining more reliable hops within the core network. This allows error-free transmission to avoid noise accumulation, while providing graceful degradation against the variations in the first hop. For the compression of the received analog DeepJSCC codeword at the first relay, a neural compression scheme is employed. Note that, when we increase the compression rate, we can achieve faster transmission over the core network, at the expense of some loss in the reconstruction quality. To satisfy different user requirements, e.g., latency and reconstruction quality, different JSC models are trained, balancing the reconstruction quality with the latency over the core network. We further show that the proposed JSC framework avoids the cliff and leveling effects where the neural compressor at the first relay can judiciously generate bit streams even if there is a mismatch between the experienced channel quality and the training value. While JSC performs slightly worse than DeepJSCC in a single-hop channel \cite{relay_jscc}, its benefits become clear as the number of hops increases. We would like to highlight that, while we model all the links in the network as additive white Gaussian noise (AWGN) channels, our solution will apply to any scenario where the first AWGN hop is followed by error-free finite-rate links in the core network.

\begin{figure}[!t]
\centering
\includegraphics[width=0.8\columnwidth]{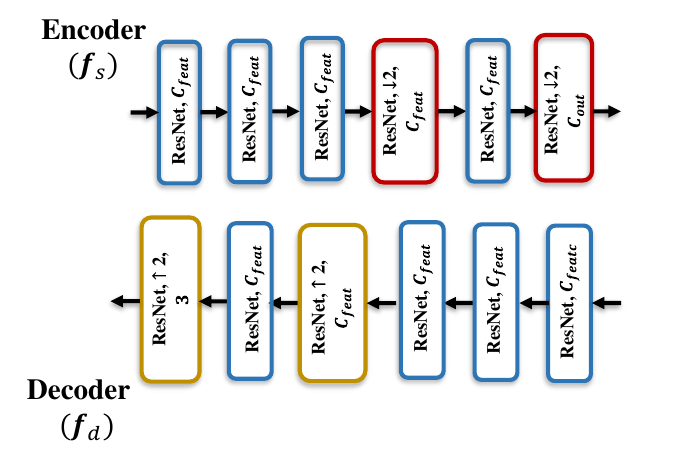}
\caption{Neural network architecture for the proposed DeepJSCC encoder $f_s(\cdot)$ and decoder $f_d(\cdot)$.}
\label{fig:fig_NN}
\end{figure}

\section{Problem Formulation}\label{sec:problem}

\subsection{System Model}\label{sec:system_model}
We consider a multi-hop network, where a mobile edge user $\mathrm{S}$ aims to deliver an image to a destination node $\mathrm{D}$ with the aid of $(n-1)$ relay nodes $\{\mathrm{R}_1, \cdots, \mathrm{R}_{n-1}\}$ residing in the mobile core network, as depicted in Fig.~\ref{fig:fig_system}.
In our framework, we consider the channels between adjacent nodes to be AWGN channels. 
In particular, since the relay nodes and the destination node lie within the mobile core network, the links among them are more reliable than that between $\mathrm{S}$ and $\mathrm{R}_1$  in the edge network.
This framework can be easily extended to slow-fading channels when the channel state information (CSI) is known to all nodes in the network.

The edge user $\mathrm{S}$ encodes an image $\bm{S} \in \mathbb{R}^{C\times H \times W}$ to a latent vector $\bm{x}_s \in \mathbb{C}^k$ using an encoding function $f_s(\cdot): \mathbb{R}^{C\times H \times W} \rightarrow \mathbb{C}^k$, where $C$, $H$, $W$ denote the number of color channels, the height and width of the image, respectively. The latent vector is subject to a power constraint:
\begin{equation}
\frac{1}{k}\|\bm{x}_s \|^2 \leq 1.
\label{equ:pn}
\end{equation}

The signal received at the first relay node $\mathrm{R}_1$ (i.e., the access point of the edge network) can be expressed as
\begin{equation}
\bm{y}_{1} = \bm{x}_s + \bm{w}_{s},
\label{equ:y1}
\end{equation}
where $\bm{w}_s \sim \mathcal{CN}(0, \sigma_s^2 \bm{I}_{k})$. The SNR of the first hop is defined as $\mathrm{SNR}_s \triangleq \frac{1}{\sigma_s^2}$. 

In the mobile core network, the relays can adopt different processing strategies, such as a fully digital scheme, a fully analog scheme, and the proposed JSC scheme detailed later. In general, we denote by $f_{\mathrm{R}_i}(\cdot)$ the operations at $\mathrm{R}_i$, $\bm{x}_i$ the signal transmitted from the $i$-th relay to the $(i+1)$-th relay, and $\bm{y}_i$ the signal received by the $i$-th relay from the $(i-1)$-th relay. Thus, we have
\begin{eqnarray}
&&\hspace{-0.5cm}\bm{x}_i = f_{\mathrm{R}_i}(\bm{y}_i)\in \mathbb{C}^{k_i}, \\ 
&&\hspace{-0.5cm} \bm{y}_{i+1} = \bm{x}_{i} + \bm{w}_{i}.
\label{equ:yi}
\end{eqnarray}
where $i=1,2,...,n-1$ and $\bm{w}_{i} \sim \mathcal{CN}(0, \sigma_{i}^2 \bm{I}_{k_{i}})$.  In particular,
\begin{itemize}
    \item $\bm{x}_i$, $\forall i$, are subject to the same power constraint as \eqref{equ:pn}.
    \item $k_1, \ldots, k_{n-1}$ are determined by the latency requirement of the destination, and may not be equal.
    \item all relays work in a half-duplex mode: $\mathrm{R}_i$ starts to transmit $\bm{x}_{i}$ after it receives and processes the entire $\bm{y}_i$.
    \item the noise variance in the edge network $\sigma_s^2$ is generally larger than that within the mobile mobile core network $(\sigma^2_1,\ldots,\sigma^2_{n-1})$. 
    \item To simplify the presentation, we assume that all the link qualities in the mobile core network are identical, i.e., $\sigma^2_{1:(n-1)} = \sigma^2$, and the SNR is denoted by $\mathrm{SNR}_N = \frac{1}{\sigma^2}$.
\end{itemize}

Finally, the received signal at the destination $\mathrm{D}$ can be written as $\bm{y}_d \triangleq \bm{y}_n = \bm{x}_{n-1} + \bm{w}_{n-1}$.
Upon receiving $\bm{y}_d$, the destination reconstructs the original image by a decoding function $f_{\mathrm{D}}(\cdot)$ as $\hat{\bm{S}} = f_{\mathrm{D}}(\bm{y}_d)$. To evaluate the reconstruction performance of different processing strategies (specified by $f_s$, $\{f_{\mathrm{R}_i},\forall i\}$, $f_{\mathrm{D}}$), we adopt the peak signal-to-noise ratio (PSNR) of the image as the metric, given by
\begin{align}
    \text{PSNR} &= 10\log_{10} \frac{255^2}{\frac{1}{CHW}||\bm{S}-\hat{\bm{S}}||^2_F}.
    \label{eq:psnr}
\end{align}

\begin{figure*}[t]
\centering
\includegraphics[width=0.8\linewidth]{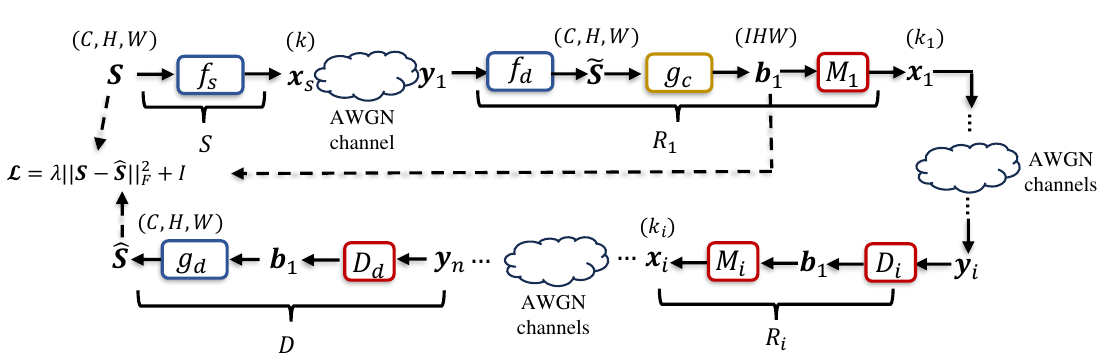}\\
\caption{Illustration of the proposed JSC framework, where $f_s(\cdot)$ denotes the DeepJSCC encoder, $f_d(\cdot)$ denotes the DeepJSCC decoder located at $\mathrm{R}_1$ and $\mathrm{D}$. $g_c(\cdot)$ and $g_d(\cdot)$ represent the compression and decompression modules, $M_i, D_i$ refer to the coded modulation block and the channel decoding block at the $i$-th relay, respectively. We show the dimensions on top of the tensors and vectors for clarity. The loss function, $\mathcal{L}$, is a summation of the distortion and the bpp ($I$) defined in \eqref{equ:Izv} used to compress the image $\widetilde{S}$.}
\label{fig:jsc_framework}
\end{figure*}

\subsection{Existing Methods}
We first review existing methods for communication over multi-hop wireless networks, namely, the fully analog and fully digital transmission.

\subsubsection{Fully analog transmission}\label{section:fat}
Transmitted signal at each hop is produced by a DeepJSCC encoder. Since the relay cannot recover the message exactly, the signal gradually becomes noisier with each hop. 
Two protocols are considered, namely, the DeepJSCC-AF and DeepJSCC-PF corresponding to the conventional amplify-and-forward and decode-and-forward schemes. We briefly introduce them as follows and refer to \cite{relay_jscc} for more details.


\textit{DeepJSCC encoder/ decoder}: As shown in Fig. \ref{fig:fig_NN}, the encoder and decoder both rely on convolutional neural networks (CNN) as their backbone and `ResNet' in the figure refers to a block of 2D CNN with residual connections. At the encoder, two downsampling layers, each with a stride of 2, are used and the numbers of hidden and output channels are denoted by $C_{feat}$ and $C_{out}$, respectively. At the decoder, two upsampling blocks, each with stride of 2, realized by pixel shuffling operations, restore the original resolution $(H, W)$. The number of output channels at the decoder is $C$.

In the DeepJSCC-AF protocol, since the relay nodes simply amplify their received signals, we have $k_1 = \cdots = k_{n-1} = k^\prime$. Note that the number of complex channel uses $k$ for the first hop may be different. We employ neural networks comprised of 2D CNNs at $\mathrm{R}_1$ to change the number of channel uses (for $k \neq k^\prime$ case) by controlling the number of output channels.
The processing at the $i$-th relay node, $\mathrm{R}_i$ can be expressed as $\bm{x}_{i} = \alpha\bm{y}_{i}$, $\quad i=2, \ldots, n-1$ with $\alpha = \sqrt{\frac{1}{1+\sigma^2}}$. 
The loss function is simply:
\begin{equation}
    \mathcal{L}_{AF} = \mathbb{E}_{\bm{S}\sim P_{\bm{S}}}\left[||\bm{S} - \hat{\bm{S} }||^2_2\right].
    \label{equ:AFLoss}
\end{equation}

In DeepJSCC-PF \cite{relay_jscc}, we instead employ neural networks to generate the signals transmitted by the relays, instead of simple amplification.  
At $\mathrm{R}_1$, DeepJSCC-PF is identical to that in DeepJSCC-AF (except for the case when $k = k^\prime$, where $\mathrm{R}_1$ of the DeepJSCC-AF protocol simply amplifies the signal), while for the remaining hops, the input-output relationship for $\mathrm{R}_i$ can be written as:
\begin{equation}
\bm{x}_{i} = f_{\mathrm{R}_i}(\bm{y}_{i}), \quad 2 \le i \le (n-1).
\label{equ:pf}
\end{equation}
where the forwarding functions $f_{\mathrm{R}_i}(\cdot)$ are parameterized by neural networks. 
DeepJSCC-PF employs the same loss function as in AF.

\subsubsection{Fully digital transmission}\label{section:fdt} The encoder at $\mathrm{S}$ first compresses the input image into a bit sequence, denoted by $\bm{b}_s$, which is then channel coded and modulated to form the channel codeword $\bm{x}_s$. Since the state-of-the-art source encoders (e.g., BPG) output variable-length codewords, the number of complex channel uses $k$ will refer to the expected length of $\bm{x}_s$. The processing operation $f_{\mathrm{R}_1}(\cdot)$ at the first relay is comprised of a channel decoder, denoted as $D_1(\cdot)$, to decode the received signal $\bm{y}_1$ to the bit sequence $\bm{b}_s$ and re-encodes it to $\bm{x}_1$ using an appropriate coded modulation scheme, $M_1(\cdot)$. Since we assume identical channels over the mobile core network, $M_i(\cdot)$ and $D_i(\cdot)$'s adopted by the relay nodes are the same.

To better illustrate the processing of the full digital transmission scheme, we provide a concrete example where $\mathrm{SNR}_s = 2$ dB and $\mathrm{SNR}_{N} = 10$ dB, 
with corresponding channel capacities $\mathcal{C}_s = 1.37$ and $\mathcal{C}_N = 3.46$, respectively. We adopt two coded modulation schemes with rates below these capacities (for reliable transmission). To be precise, we consider a rate-1/2 LDPC code with 4-QAM corresponding to a achievable rate $R_s = 1$ for the first link, and a rate-1/2 LDPC code with 16-QAM, which corresponds to $R_N = 2$, for the remaining links. It has been shown by simulations that the coded modulation schemes along with the corresponding belief propagation (BP) decoders are capable to achieve almost zero packet error rate (PER) in the considered scenario. We also note that $k = 2k^\prime$ holds for this case.

A key weakness of the fully digital transmission is that, though it is plausible to assume $\mathrm{SNR}_N$ is stable, due to the link between the source and the first relay node is versatile, it will still suffer from the cliff and leveling effects.

\section{The proposed JSC framework}\label{sec:JSC}
\subsection{Hybrid Solution}
We consider a hybrid scheme (see Fig. \ref{fig:jsc_framework}), where the mobile user adopts DeepJSCC protocol in the first hop to avoid the cliff and leveling effects, whereas the first relay node, $\mathrm{R}_1$, uses a deep neural network, $g_c(\cdot)$ to compress its received signal, $\bm{y}_1$ to a bit sequence $\bm{b}_1$. Then, each $\mathrm{R}_i$ adopts a coded modulation scheme, $M_i(\cdot)$, which is identical to the fully digital scheme in Section \ref{section:fdt} for transmission over the $(n-1)$ hops in the mobile core network.  All coded modulation schemes are carefully chosen to exhibit zero PER; thus, it can be guaranteed that $\mathrm{D}$ correctly decodes the bit sequence $\bm{b}_1$ regardless of the number of hops. This avoids performance degradation caused by noise accumulation as in the fully-analog scheme when the number of hops increases. Then, the decoder at $\mathrm{D}$, takes $\bm{b}_1$ as input to generate the final reconstruction, $\hat{\bm{S}}$. Note that the processing functions, $\{f_{\mathrm{R}_2}(\cdot), \ldots, f_{\mathrm{R}_{n-1}}(\cdot)\}$  at $\{\mathrm{R}_2, \ldots, \mathrm{R}_{n-1}\}$ only involve channel decoding $D_i(\cdot)$ and re-encoding $M_i(\cdot)$ without neural processing.

\subsubsection{Encoder at $\mathrm{S}$} 
The encoder function $f_s(\cdot)$ at $\mathrm{S}$ is shown in Fig. \ref{fig:fig_NN}, and is identical to the DeepJSCC encoder introduced in Section \ref{section:fat}. 

\subsubsection{Processing at $\mathrm{R}_1$}
A naive solution is to directly quantize each element of the received signal $\bm{y}_1$ to a given precision, e.g., $m$ bits, and protect the bit sequence with length $km$ using a coded modulation scheme. However, naively quantizing the received signal leads to sub-optimal solution, which motivates the employment of a neural compression model. 

Instead of directly compressing $\bm{y}_1$, we found it is more beneficial to first transform $\bm{y}_1$ into a tensor with the same dimension of the original image $\bm{S}$, denoted by $\widetilde{\bm{S}}$,  via a DeepJSCC decoder, $f_d(\cdot)$, introduced in Section \ref{section:fat}, and then apply the neural image compression algorithm \cite{image_compress2}, denoted as $g_c(\cdot)$, to compress $\widetilde{\bm{S}}$ into a bit sequence $\bm{b}_1$. The processing of the neural compression model is summarized as follows (see also Fig. \ref{fig:jsc_framework}): After obtaining the  reconstructed image $\widetilde{\bm{S}}$, we first use non-linear transform blocks, denoted by $g_a(\cdot)$, which is comprised of 2D CNNs with $C_z$ output channels to generate a latent tensor $\bm{z}\in \mathbb{R}^{C_z\times H/4 \times W/4}$, which is fed to the hyper-latent encoder, $h_a(\cdot)$, to generate the hyper-latent $\bm{v}\in \mathbb{R}^{C_v\times H/16 \times W/16}$. The hyper-latent $\bm{v}$ is introduced with the aim to remove the dependency between the elements in $\bm{z}$, such that given $\bm{v}$, the probability density function (pdf) of the elements in $\bm{z}$ will be in product form. 

It is worth mentioning that the neural image compression model involves two phases, the training and deployment phases. In the deployment phase, we adopt arithmetic encoder to encode both latent vectors, where each element, ${z}_i$ and ${v}_i$, are rounded to the nearest integers, $\hat{{z}}_i$ and $\hat{{v}}_i$, and then compressed according to their individual learned distributions. In the training phase, however, the rounding operation is infeasible as its gradient is one only at integer points, and zero otherwise. Thus, during training, we use $\tilde{\bm{z}}, \tilde{\bm{v}}$ to replace the rounded vectors $\hat{\bm{z}}, \hat{\bm{v}}$, where each element $\tilde{{z}}_i, \tilde{{v}}_i$ is obtained by adding uniform noise (to model the quantization noise) to ${z}_i, {v}_i$. A hyper latent decoder, $h_s(\cdot)$, with two upsampling layers takes $\tilde{\bm{v}}$ as input, and outputs two tensors $\tilde{\bm{\mu}}$ and $\tilde{\bm{\sigma}}$ with the same dimensions as $\tilde{\bm{z}}$. Then, we model the pdf of $\tilde{\bm{z}}$ as:
\begin{align}
    p(\tilde{\bm{z}}|\tilde{\bm{v}}) = &\prod_i \left(\mathcal{N}(\tilde{\mu}_i, \tilde{\sigma}^2_i)*\mathcal{U}(-\frac{1}{2}, \frac{1}{2}) \right)(\tilde{z}_i), \notag \\
    \widetilde{\bm{\mu}}, \widetilde{\bm{\sigma}} &= h_s(\tilde{\bm{v}}, \bm{\theta}),
\label{equ:probz}
\end{align}
where $\bm{\theta}$ denotes the parameters for $h_s(\cdot)$ and $*$ represents convolution.
For the hyper latent $\tilde{\bm{v}}$, we consider modeling it also using a fully factorized model as:
\begin{align}
    p(\tilde{\bm{v}}|\bm{\phi}) = \prod_i \left(p_{\tilde{v}_i|\phi_i}(\phi_i)*\mathcal{U}(-\frac{1}{2}, \frac{1}{2}) \right)(\tilde{v}_i),
\label{equ:probv}
\end{align}
where $\bm{\phi}$ denotes a collection of vectors to parameterize the univariate distribution. 

\begin{figure}[!t]
\centering
\includegraphics[width=0.8\columnwidth]{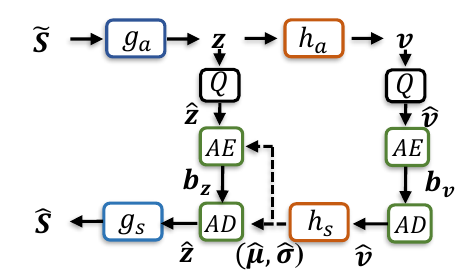}
\caption{Illustration of the compression and decompression modules, $g_c(\cdot)$ and $g_d(\cdot)$, respectively. In particular, $g_a(\cdot), h_a(\cdot)$ are the non-linear transform blocks to generate the latent vectors $\bm{z, v}$, whereas $g_s(\cdot), h_s(\cdot)$ are adopted to facilitate the synthesis of $\hat{\bm{S}}, \hat{\bm{z}}$. The predicted tensors $\hat{\bm{\mu}}, \hat{\bm{\sigma}}$ are used for the arithmetic encoding and decoding of $\hat{\bm{z}}$ and $\bm{b}_z$.}
\label{fig:fig_hyperprior}
\end{figure}

Once the models are trained and deployed, the latent and hyper-latent vectors $\bm{z}, \bm{v}$ can be obtained from the reconstructed image $\widetilde{\bm{S}}$. We use an arithmetic encoder to encode the rounded $\hat{\bm{z}}, \hat{\bm{v}}$ to the bit sequence $\bm{b}_1 = (\bm{b}_z, \bm{b}_v)$ according to the optimized pdfs in \eqref{equ:probz} and \eqref{equ:probv} by simply replacing $\tilde{\bm{z}}, \tilde{\bm{v}}$ by $\hat{\bm{z}}, \hat{\bm{v}}$. The bit sequence will be further coded and modulated before forwarding to $\mathrm{R}_2$. Due to the fact that different images contain different amount of information, the length of the bit sequence varies and we evaluate the expected number of complex channel uses as 
\begin{equation}
    k^\prime = 
\frac{1}{R_N} \mathbb{E}_{\widetilde{\bm{S}} \sim p_{\widetilde{\bm{S}}}}(\#\bm{b}_1),
\label{equ:k_prime}
\end{equation}
where $\#\bm{b}_1$ denotes the number of bits in the sequence while $R_N$ is defined in Section \ref{section:fdt}. 

\subsubsection{Neural processing at $\mathrm{D}$} 
Since the links in the mobile core network are assumed static, and protected by channel codes exhibiting zero PER, the destination node can retrieve the bit sequence $\bm{b}_1$ correctly, and employs an arithmetic decoder ($AD$) to first generate the rounded latent vector $\bm{\hat{v}}$, which is then fed to $h_s(\cdot)$ to obtain $\hat{\bm{\mu}}, \hat{\bm{\sigma}}$. Then, another $AD$ takes the mean and variance, $\hat{\bm{\mu}}, \hat{\bm{\sigma}}$ along with the sequence $\bm{b}_z$ to produce the rounded latent vector $\bm{\hat{z}}$. Finally, the DeepJSCC decoder, $f_d(\cdot)$ takes $\bm{\hat{z}}$ as input and outputs $\hat{\bm{S}}$. 

\textbf{Loss function for the JSC framework}.
Similarly to the general source compression problem, there is a trade-off between the amount of bits used to compress $\bm{y}_1$ (or equivalently, $\bm{\widetilde{S}}$) and the reconstruction performance at $\mathrm{D}$. To achieve different trade-off points in the rate-distortion plane, we introduce a variable $\lambda$ and the loss function is written as:
\begin{equation}
    \mathcal{L} = \lambda \|\bm{S} - \hat{\bm{S}}\|^2_F + I,
\end{equation}
where $I = I_z + I_v$ is the summation of the bit per pixel (bpp) for compressing $\tilde{\bm{z}}$ and $\tilde{\bm{v}}$, which is defined by the number of bits divided by the height $H$ and width $W$ of the image. Note that the notation $\tilde{\bm{z}}, \tilde{\bm{v}}$ indicate the training phase\footnote{Definition of $I$ is the same for the deployment phase obtained by replacing $\tilde{\bm{z}}$ and $\tilde{\bm{v}}$ with $\hat{\bm{z}}$ and $\hat{\bm{v}}$.} and the rate can be further expressed as:
\begin{align}
    I = \frac{1}{HW} \mathbb{E}_{\tilde{\bm{z}}, \tilde{\bm{v}} \sim q} \left[-\log_2(p_{\tilde{\bm{z}}|\tilde{\bm{v}}}(\tilde{\bm{z}}|\tilde{\bm{v}})) - \log_2(p_{\tilde{\bm{v}}}(\tilde{\bm{v}})) \right],
\label{equ:Izv}
\end{align}
where the first and second terms represent $I_z$ and $I_v$, respectively, while $q$ denotes the posterior of $\tilde{\bm{z}}, \tilde{\bm{v}}$ given reconstructed image $\widetilde{\bm{S}}$ at $\mathrm{R}_1$, which follows a uniform distribution centered at $g_a(\cdot)$ and $h_a(\cdot)$ output, ${\bm{z}}$ and ${\bm{v}}$:
\begin{align}
    q(\tilde{\bm{z}}, \tilde{\bm{v}}|\bm{\widetilde{S}}) = \prod_i \mathcal{U}(\tilde{{z}}_i|{z}_i-\frac{1}{2}, {z}_i+\frac{1}{2}) \cdot \mathcal{U}(\tilde{{v}}_i|{v}_i-\frac{1}{2}, {v}_i+\frac{1}{2}).
\label{equ:post}
\end{align}

\section{Numerical Experiments}

\label{sec:experiment}
\subsection{Parameter Settings and Training Details}
Unless otherwise mentioned, the number of complex channel uses, $k$ for the first hop is fixed to 768, which corresponds to $C_{out}=24$. The number of channels, $C_z$ and $C_v$, for $\bm{z}$ and $\bm{v}$ are set to 256 and 192, respectively.

Similar to the training setting in \cite{relay_jscc}, we compare the proposed JSC framework with the fully digital and fully analog schemes, considering the transmission of images from the CIFAR-10 dataset, which consists of $50,000$ training and $10,000$ test images with $32 \times 32$ resolution. Adam optimizer is adopted with a varying learning rate, initialized to $10^{-4}$ and dropped by a factor of $0.8$ if the validation loss does not improve in $10$ consecutive epochs.

\subsection{Performance Evaluation}

\begin{figure}[!t]
\centering
\includegraphics[width=0.8\columnwidth]{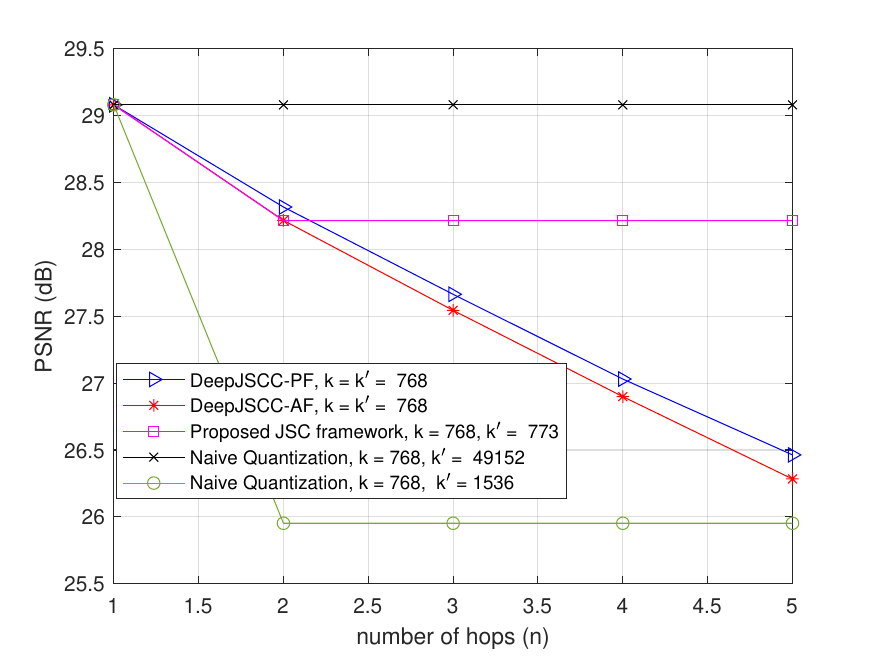}
\caption{The comparison of PSNR performance between the proposed JSC framework along with the DeepJSCC-PF and DeepJSCC-AF protocol. We set $k =k^\prime = 768$ for the two benchmarks, while $\lambda = 3200, k^\prime = 773$ for the proposed framework.}
\label{fig:fig_jsc_performance}
\end{figure}

\begin{figure}[!t]
\centering
\includegraphics[width=0.8\columnwidth]{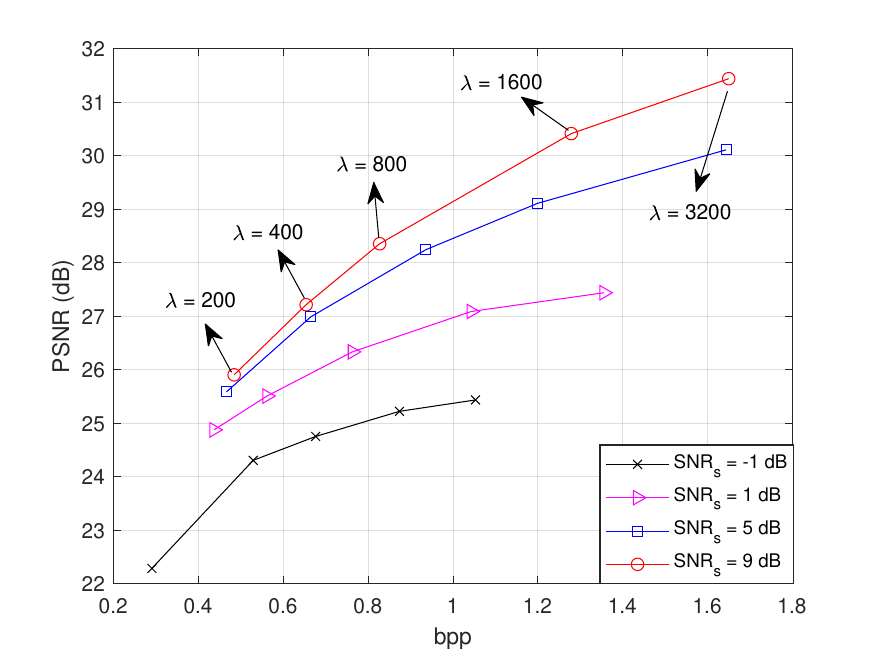}
\caption{Rate distortion tradeoff for the proposed JSC framework where 20 JSC models are trained and evaluated corresponding to $SNR_s \in \{-1, 1, 5, 9\}$ dB and $\lambda \in \{200, 400, 800, 1600, 3200\}$.}
\label{fig:fig_rdtradeoff}
\end{figure}

\subsubsection{Comparison with fully analog and fully digital schemes}
We consider the scenario presented in Section \ref{sec:problem}, where the SNR between the mobile user $\mathrm{S}$ and the first relay $\mathrm{R}_1$ is $\mathrm{SNR}_s = 2$ dB, while the link qualities within the core network are $\mathrm{SNR}_N = 10$ dB. 

We first compare the performance between the proposed JSC framework with fully analog transmission using DeepJSCC-AF and DeepJSCC-PF, respectively. We make simplification such that $k^\prime = k$, thus, for DeepJSCC-AF, all relays only need to amplify the signal and the effective SNR, $\mathrm{SNR}_{eff}$ at $\mathrm{D}$ is expressed as:
\begin{align}
    \mathrm{SNR}_{eff} = \frac{1}{\sum_{i=0}^{n-1} \sigma^2_i \prod_{j=0}^{i-1}(1 + \sigma^2_j)},
    \label{equ:eff_snr}
\end{align}
where we use the notation $\sigma^2_0$ to replace $\sigma^2_s$.

Since the coded modulation adopted in this case is a 1/2-rate LDPC with 16-QAM, for a fair comparison, the bpp, $I$, used to compress the received signal $\bm{y}_1$ (or equivalently, $\bm{\widetilde{S}}$) should satisfy $I \approx \frac{2k^\prime}{HW} = 1.5$ averaged over all images from the dataset. To achieve this, we carefully select the hyper-parameter, $\lambda = 3200$ leading to $I = 1.52$. Fig. \ref{fig:fig_jsc_performance} shows the relative performance of the proposed JSC framework, the naive quantization solution, and the DeepJSCC-AF/PF protocols. When the number of hops is one, i.e., a point-to-point channel, all schemes yields the same PSNR performance around $29.1$ dB. When there is at least one relay in the network, due to the lossy compression at $\mathrm{R}_1$ for the proposed JSC scheme, its performance drops to $\approx 28.2$ dB and then keeps constant since the noise introduced by each hop is removed by the channel code before being forwarded to the next relay; that is, there is no noise accumulation. The DeepJSCC-AF and PF protocols, however, both suffer from noise accumulation, and their performances drop as the number of hops increases. To be precise, when the number of hops becomes larger than 3, both of them are outperformed by the proposed JSC framework. We also observe that DeepJSCC-PF outperforms DeepJSCC-AF only by a small margin, although requiring the training of neural network processors at all the relays.
The curves marked `Naive Quantization' represent the scheme, in which each element of $\bm{y}_1$ is quantized to 4 and 128 bits\footnote{Note that we consider a non-uniform scalar quantization for the 4-bit quantization shown in the figure where the codebook is obtained via Lloyd algorithm. We also consider 64-bit quantization as a bound since the precision of complex numbers in Pytorch is 128 bits.} and further protected by channel codes. It can be seen that the proposed scheme outperforms the 4-bit quantization baseline with smaller $k^\prime$. The $128$-bit precision scheme, as expected, outperforms all other schemes, but it only serves as an upper bound with no loss over the hops, as the overhead, i.e., $k^\prime$, makes it unpractical.

We then compare the proposed scheme with the fully digital scheme. In this case, the digital scheme uses a 1/2-rate LDPC with 4-QAM for the first hop leading to an expected number of compressed bits $\# \bm{b}_1 = k = 768$. The coded modulation scheme for the remaining relays has a rate $R_N = 2$, thus we have $k^\prime = k/2$. For a fair comparison, the number of compressed bits for the JSC framework should also be less or equal to $k$. We choose $\lambda = 800$ with a corresponding $\# \bm{b}_1 = 646$ and find that the proposed scheme achieves $\text{PSNR} =26.19$ dB, which is higher than $25.39$ dB achieved by the fully digital scheme.

\subsubsection{Rate-distortion curves for the proposed JSC framework} 
Next, we focus on the proposed JSC framework and show that by training models with different $\lambda$'s, the JSC framework can achieve different points on the rate-distortion plane, enabling flexible adaptation to the user specific reconstruction requirement and latency constraint. In this simulation, we do not specify the coded modulation scheme used within the core network but assume the compressed bits generated by $\mathrm{R}_1$ can be delivered to the destination without error. Different $\mathrm{SNR}_s \in \{-1, 1, 5, 9\}$ dB  are evaluated and for each of the SNR value, we train different models with respect to different $\lambda \in \{200, 400, 800, 1600, 3200\}$ values. 

We plot the PSNR performance versus bpp $I$ defined in \eqref{equ:Izv} in Fig. \ref{fig:fig_rdtradeoff}. For all $\mathrm{SNR}_s$ values, different bpp values are obtained with different models. When the latency constraint of the destination is stringent, we adopt a model with small $\lambda$, when the destination requires a better reconstruction performance, however, a large $\lambda$ is preferable. We also find a better rate-distortion curve is obtained when $\mathrm{SNR}_s$ is larger, which is intuitive as the first hop becomes less noisy. 

\begin{figure}[!t]
\centering
\includegraphics[width=0.8\columnwidth]{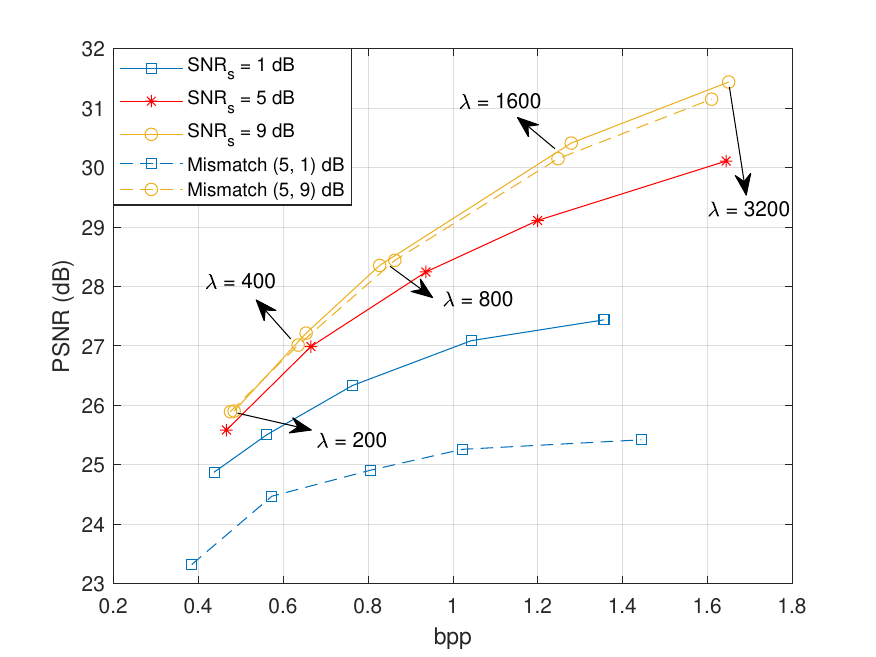}
\caption{Different models trained at $\mathrm{SNR}_s = 5$ dB with $\lambda \in \{200, 400, 800, 1600, 3200\}$ are evaluated under $\mathrm{SNR}_s = \{1, 9\}$ dB.}
\label{fig:fig_cliffeffect}
\end{figure}

\subsubsection{Avoiding the cliff and leveling effects} 
Finally, we show the proposed JSC framework is capable of avoiding the cliff and leveling effects. In this simulation, we consider training different models with different $\mathrm{SNR}_s\in \{1, 5, 9\}$ dB and the aforementioned $\lambda$ values. Then, we evaluate the performance of the five models corresponding to five different $\lambda$ values trained at $\mathrm{SNR}_s = 5$ dB while tested at SNR values $\{1, 9\}$ dB. As shown in Fig. \ref{fig:fig_cliffeffect}, the proposed JSC framework has no cliff or leveling effects as a better rate-distortion curve is obtained when tested at higher SNR while a reasonable performance is maintained when the SNR drops below the SNR trained. 
This verifies the effectiveness of the proposed JSC framework.

\section{Conclusion}
In this paper, a novel JSC framework is proposed to facilitate wireless image transmission over multi-hop networks. In particular, a mobile user delivers the message to its destination node with the aid of the relays, where all the nodes expect the mobile user are located in a mobile core network with stable wireless connections.
It is shown that naively applying fully digital or fully analog transmissions will lead to unsatisfactory results. Thus, we propose a hybrid solution where DeepJSCC is adopted for the first hop and a neural compression model is employed at the first relay node to convert the DeepJSCC codeword into bits for transmission over the mobile core network. Simulation results show that the proposed JSC framework is able to outperform both the fully analog and fully digital schemes, and provides robustness against variations in the first wireless hop, while retaining low delay over the mobile core network.

\bibliographystyle{IEEEbib}
\bibliography{refs}


\end{document}